\documentclass[]{aa}  
\usepackage{graphicx}
\usepackage{txfonts}
\usepackage{amsmath}
\usepackage{amssymb}
\usepackage{url}
\usepackage{times}
\usepackage[utf8]{inputenc}
\usepackage{xspace}
\usepackage{color}
\usepackage{multirow}
\usepackage{multicol}
\usepackage{float}
\usepackage{verbatim}
\usepackage{mathtools, cuted}
\usepackage[normalem]{ulem}
\usepackage{soul}
\usepackage{booktabs}
\usepackage[dvipsnames]{xcolor}
\usepackage{bm}
\usepackage{appendix}

\bibpunct{(}{)}{;}{a}{}{,}

\usepackage{etoolbox}
\makeatletter
\patchcmd\@combinedblfloats{\box\@outputbox}{\unvbox\@outputbox}{}{%
  \errmessage{\noexpand\@combinedblfloats could not be patched}%
}%
\makeatother

\def\msun{\,\rm M_\odot}

\begin{document} 

\title{\textcolor{black}{Distinguishing the nature of “ambiguous'' merging systems hosting a neutron star: GW190425 in low-latency}}

\author{C. Barbieri
          \inst{\ref{unimib},\ref{infn.mib},\ref{oab.me}\thanks{c.barbieri@campus.unimib.it}
        }, O.~S.~Salafia\inst{\ref{oab.me},\ref{infn.mib}},  M.~Colpi\inst{\ref{unimib},\ref{infn.mib}}, G.~Ghirlanda\inst{\ref{oab.me},\ref{infn.mib}} and
        A.~Perego\inst{\ref{unitn},\ref{TIFPA}}}

\institute{Università degli Studi di Milano-Bicocca, Dipartimento di Fisica ``G. Occhialini'', Piazza della Scienza 3, I-20126 Milano, Italy\label{unimib}\and INAF -- Osservatorio Astronomico di Brera, via E. Bianchi 46, I-23807 Merate, Italy\label{oab.me}  \and INFN -- Sezione di Milano-Bicocca, Piazza della Scienza 3, I-20126 Milano, Italy\label{infn.mib} \and Universit\`a degli Studi di Trento, Dipartimento di Fisica, via Sommarive 14, I-38123 Trento, Italy \label{unitn}
\and INFN-TIFPA, Trento Institute for Fundamental Physics and Applications, via Sommarive 14, I-38123 Trento, Italy \label{TIFPA}
}

\authorrunning{C.~Barbieri  et. al}

\date{Received XXX; accepted XXX}

\abstract{\textcolor{black}{GW190425 is the newly discovered gravitational wave (GW) source consistent with a neutron star-neutron star merger with chirp mass of $1.44\pm0.02\msun.$ This value falls in the \emph{ambiguous} interval as from the GW signal alone we can not rule out the presence of a black hole in the binary. In this case, the system would host a neutron star and a very light stellar black hole, with mass close to the maximum value for neutron stars, filling the \emph{mass gap}. No electromagnetic counterpart is firmly associated with this event, due to the poorly informative sky localisation and larger distance, compared to GW/GRB170817.
We construct here kilonova light curve models for GW190425, for both double neutron star and black hole-neutron star systems, considering two equations of state consistent with current constraints from the signals of GW170817/GW190425 and the NICER results, including black hole spin effects and assuming a new formula for the mass of the ejecta. The putative presence of a light black hole in GW190425 would have produced a brighter kilonova emission compared to the double neutron star case, letting  us to distinguish the nature of the companion to the neutron star. Concerning candidate counterparts of GW190425, classified later on as supernovae, our models could have discarded two transients detected in their early \textit{r}-band evolution. Combining the chirp mass and luminosity distance information from the GW signal with a library of kilonova light curves helps identifying the electromagnetic counterpart early on. We remark that the release in low latency  of the chirp mass in this interval of \emph{ambiguous} values appears to be vital for successful electromagnetic follow-ups.}}


\maketitle

\section{Introduction}\label{sec:intro}
The LIGO Scientific Collaboration and Virgo Collaboration (LVC) detected gravitational waves (GWs) from the inspiral and merger of several stellar origin black hole-black hole (BHBH) binaries \citep{Abbott18-10-bh}, during the observing runs O1 and O2 (2015-2017). In August 2017, the first neutron star-neutron star (NSNS) binary coalescence was detected \citep[GW170817,][]{GW170817}, which was accompanied by broad-band electromagnetic (EM) counterparts \citep{gw170817em}, heralding the birth of the multi-messenger GW-EM astronomy. Recently,
during the O3 run, the second NSNS merger was detected \citep[GW190425, ][]{GW190425}, but no EM counterpart was firmly associated with this event\footnote{\textcolor{black}{\citealt{Pozanenko2019} suggest an association with GRB190425, although \cite{Foley2020,Song2019} indicate that data are not constraining.}} \citep{Coughlin2019_2}.

The merger of a black hole-neutron star (BHNS) binary represents an highly anticipated GW source \citep{Abadie2010}. At the time of this writing, LVC reported promising candidates\footnote{A complete list of candidates is available on the LIGO/Virgo O3 Public Alerts webpage \url{https://gracedb.ligo.org/superevents/public/O3/}.}, such as S190814bv \citep{S190814bv} and S190910d \citep{S190910d}. No EM counterpart was associated with these candidates \citep[see][and references therein]{Coughlin2019_3}.

It is anticipated that BHNS mergers can produce EM counterparts as NSNS mergers do, mainly depending on the combination of four binary parameters, namely the BH mass $M_\mathrm{BH}$ and spin\footnote{Hereafter, $\chi_\mathrm{BH}=cJ/GM_\mathrm{BH}^2$ is the dimensionless spin parameter and $J$ is the BH angular momentum.} $\chi_\mathrm{BH}$, the NS mass $M_\mathrm{NS}$ and tidal deformability $\Lambda_\mathrm{NS}$. The latter depends on the equation of state (EoS) of NS matter \citep{Shibata11,Foucart2012,Kyutoku2015,Kawaguchi2015,Foucart2018}. In particular the optimal condition to favor NS tidal disruption, and therefore the ejecta release that powers EM counterpart emission, is to have low mass ratio $q=M_\mathrm{BH}/M_\mathrm{NS}$, large $\chi_\mathrm{BH}$ and large $\Lambda_\mathrm{NS}$ or, equivalently, “stiff'' EoS \citep{Bildsten1992,Shibata2009,Foucart2013,Foucart2013b,Kawaguchi2015,Pannerale2015a,Pannerale15b,Hinderer2016,Kumar2017, Barbieri2019}. At leading-order, the orbital evolution of a compact binary system is governed by a combination of the two objects masses, known as chirp mass, 
\begin{equation}
   M_\mathrm{c} = \frac{(M_1M_2)^{3/5}}{(M_1+M_2)^{1/5}}. 
\end{equation}

\textcolor{black}{\cite{Barbieri2019_3} pointed out that systems with low chirp masses, in the range $1.2\msun\lesssim M_\mathrm{c}\lesssim2\msun$ depending on the EoS, can host a larger variety of configurations. They can either be NSNS or BHNS binaries}\footnote{In this work we assume that the NS and BH mass distributions are adjacent (no “mass gap'', see discussion in \S\ref{sec:BH_in_GW190425}).}, and their nature can not be distinguished through the GW signal detection alone, at least in low-latency analysis \citep{Mandel2015}. \textcolor{black}{We defined these systems as \emph{ambiguous}.}

{\textcolor{black}{\cite{Hinderer2019} first presented a direct comparison of GW and EM observables from BHNS and NSNS mergers with same mass ratio. \cite{Kawaguchi2019} showed that BHNS mergers can produce kilonova emission as bright as the GW170817 case in the optical band and even brighter in the infrared (1–2 mag). They indicated that the different properties of the ejecta are imprinted in the different values of the peak brightness and time, suggesting that multi-wavelength kilonova observation can unveil the central engine. However \cite{Hinderer2019} considered NSNS/BHNS systems with only low-mass NSs ($M_\mathrm{NS}=1.2\msun$ and $1.44\msun$) with mass ratio $q=1$ and $q=1.2$, while \cite{Kawaguchi2019} considered BHNS systems with NS mass $M_\mathrm{NS}=1.35\msun$ and $q=3$ and $q=7$ \citep[simulations presented in][]{Kyutoku2015}, outside the critical interval of \emph{ambiguous} systems.}}

\textcolor{black}{In \cite{Barbieri2019_3} we showed that the kilonova emission from \emph{ambiguous} NSNS and BHNS mergers, corresponding to the same $M_\mathrm{c},$ can be very different due to the difference in the properties of their discs and ejecta. NSNS binaries with \emph{ambiguous} chirp masses host either a NS with $M_\mathrm{NS}\sim1.4$ and a very massive NS ($\sim2\msun$, close to the maximum allowed value $M_\mathrm{NS}^\mathrm{max}$), or two NSs with $\sim1.6-1.8\msun$. In the latter case, the mergers of massive, symmetric and low-$\Lambda_\mathrm{NS}$ stars produce very few ejecta \citep[see Fig. 28 and Fig. 2 of][respectively]{Radice2018,Barbieri2019_3} and the kilonovae from these systems can be very dim.
The variety of configurations consistent with the same \emph{ambiguous} chirp mass
gives rise to a potentially large suite of possible kilonova light curves, and in \cite{Barbieri2019_3} we showed that BHNS binaries are optimal systems for ejecta production. They can be accompanied by bright and distinguishable kilonovae despite degeneracies induced by the large set of physical parameters associated with a given system \citep[see Fig. 4 in ][]{Barbieri2019_3}. Comparisons between BHNS and NSNS mergers with mass ratio $q\sim2$ and NSs close to the maximum mass,  corresponding to \emph{ambiguous} systems, are lacking and this is a motivation to study these systems in more detail in this paper.}

\textcolor{black}{
 Distinguishing the nature of the compact object, companion to the NS, in these systems is of great value  since we can (i) narrow down the uncertainties on the maximum mass of NSs $M_\mathrm{NS}^\mathrm{max}$; (ii) discover the existence of BHs close to the maximum NS mass, and thus the absence of the compact objects “mass-gap'' between $\sim2\msun$ and $\sim5\msun$. Both results would be of paramount importance to constrain the NS EoS.}
 
\textcolor{black}{The recently discovered GW190425 is an \emph{ambiguous} binary with $M_\mathrm{c}=1.44\pm0.02$, 
for which the presence of a BH (or even two BHs) can not be completely ruled out \citep{GW190425,Kyutoku2020,Han2020}. The localization of the source was poor as only a single detector (LIGO Livingston) detected the signal with high confidence. This prevented any triangulation using time delays among interferometers.} 

\textcolor{black}{This work is an application of the study presented in \cite{Barbieri2019_3} to the real case of GW190425, whose chirp mass falls exactly in the \emph{ambiguous} range, employing both the results from low-latency GW signal analysis and the information from the EM follow-up. Moreover we updated our model assuming two physically motivated EoS and a new fitting formula for the mass of the accretion disc produced in NSNS mergers (see below). Using the information of the chirp mass only, we generate a library of kilonova light curves for GW190425, with the aim at verifying whether the detection of a transient in the $g$, $r$, and $J$ bands would have let us distinguish, early in the evolution of the transient, the nature of the compact object companion to the neutron star, considering the richness of initial configurations compatible with the measured chirp mass and the uncertainties in the EoS.} 

\textcolor{black}{This work is also motivated by the Neutron Star Interior Composition Explorer (NICER) results \citep{Miller2019,Riley2019} which provided complementary indications on the NS EoS from high precision studies of the millisecond pulsar PSR J0030+0451 \citep{Becker2000} (see \S~\ref{sec:BH_in_GW190425}). In addition to their potential mass ratio asymmetry, \emph{ambiguous} systems are expected
to modify the mass in the ejecta and we account for new numerical findings of unequal mass NSNS binaries to calculate the ejecta properties (described in Appendix \ref{ap:fit_Om}).
The difference in the expected disc mass causes different masses of ejecta arising as disc outflows, affecting the kilonova light curves.}

\textcolor{black}{The different system configurations corresponding to the same chirp mass, labeled by the components' masses (and BH spin for BHNS case), generate an envelope of light curves. In this work we explore how the light curves from the different configurations are distributed in the magnitude-time domain. We further deepen our analysis considering different sets of model parameters, among them ejecta geometry and gray opacity,
to better quantify the level of overlap among the NSNS and BHNS light curves. We also consider BHNS configurations hosting a BH lighter than the maximum mass of NSs to explore whether light curves from these systems carry any distinctive signature. Moreover we also repeat our analysis on the posterior samples from GW190425 high-latency parameter estimation, in order to compare the light curves degeneracy level and capability of distinguishing the merging system's nature between the low-latency and high-latency cases.}

\textcolor{black}{Finally, we propose a method to prioritize the follow-up of electromagnetic transients with the aim of increasing the chance probability of merger's EM counterpart detection using the expected kilonova ranges obtained with our model. We apply this method to low-latency follow-up of GW190425, as it only requires the knowledge of the chirp mass and luminosity distance (both available in GW signal low-latency analysis). We note that \cite{Margalit2019} indicated that a rapid release of GW events' chirp mass estimates to the scientific community could help optimizing the EM follow-up. \citealt{Biscoveanu2019} also found that the systematics due to low-latency search assumptions do not affect the organisation of NSNS candidates EM follow-up based on the chirp mass.}

\textcolor{black}{The paper is organised as follows. In \S~\ref{sec:BH_in_GW190425} we discuss whether GW190425 hosts a NSNS or a BHNS merger. In \S~\ref{sec:ejecta} we estimate the mass loss in  NSNS and BHNS binary systems consistent with the chirp mass of GW190425. In \S~\ref{sec:kilonova} we create a library of light curves and compute peak magnitudes for such systems.  In \S~\ref{ap:config} we show how light curves from different configurations are distributed in the magnitude-time domain. In \S~\ref{ap:variation} we study the overlap in the BHNS and NSNS light curves changing a number of model parameters. In \S~\ref{ap:ej_kn_bh_mass_ns} we study the case where BHs have masses below the maximum mass for NSs. In \S~\ref{ap:kn_ps} we apply our analysis on the posterior samples from high-latency GW190425 parameter estimation. Finally in \S~\ref{sec:chirp} we discuss how knowledge of the chirp mass 
is critical to plan concurrent EM follow-up campaigns. In particular we apply our argument to the EM follow-ups of GW190425.}

\section{A black hole in GW190425?}
\label{sec:BH_in_GW190425}

\begin{figure}
    \centering
    \includegraphics{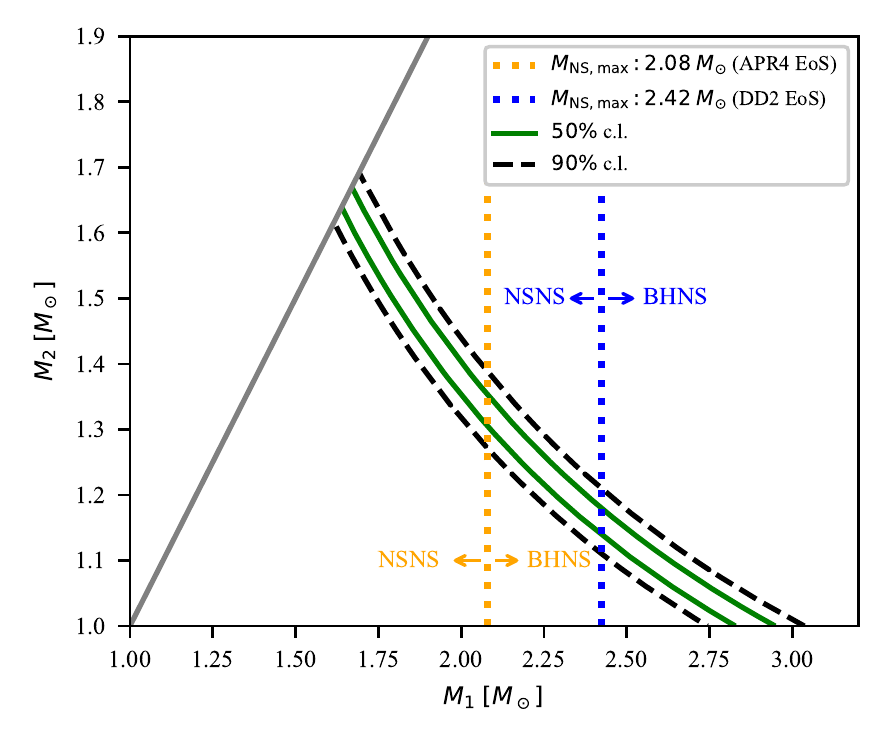}
    \caption{$M_1-M_2$ configurations compatible with the inferred value of the chirp mass for GW190425, $M_\mathrm{c}=1.44\pm0.02\msun$. We show the 50\% and 90\% confidence regions in green and dashed-black, respectively. \textcolor{black}{Orange (blue) vertical dotted line indicates the maximum NS mass for APR4 (DD2) EoS.}}
    \label{fig:chirp}
\end{figure}

\cite{GW190425} recently reported the detection of the compact object binary merger GW190425, whose chirp mass is $1.44\pm0.02\msun$. This event is most likely identified as a NSNS merger. The masses of the primary ($M_1$) and secondary ($M_2$) star are found to be in the range $1.62\msun-1.88\msun$ and $1.45\msun-1.69\msun$ (within the 90\% credible interval), respectively, assuming low-spin prior ($\chi$<0.05). Instead, assuming a high-spin prior ($\chi<0.89$), $M_1$ and $M_2$ are $1.61\msun-2.52\msun$ and $1.12\msun-1.68\msun$ (90\% credible intervals), respectively. In the case of GW190425, the poorly constrained spins and the uncertainty on the EoS that describes the NS component prevent us from clearly distinguishing a NSNS from a BHNS merger based solely on the GW signal. Therefore, the presence of a BH in GW190425 can not be excluded \textcolor{black}{\citep{GW190425,Kyutoku2020,Han2020}}, and this is possible only if there exist stellar black holes with a mass just above the maximum mass of a NS. 
The mass interval from $\sim2\msun$ to $\sim5\msun$ is usually defined as the “mass gap'', and as of today EM observations do not show evidence of BH in this mass interval \citep{Ozel2010,Farr2011}. The most massive NS is J0740+6620, with a best measure mass $M=2.14^{+0.10}_{-0.09}\msun$, while the lightest BHs detected by LVC and observed in Galactic X-ray binaries have a mass $7.6^{+1.3}_{-2.1}\msun$ \citep{Abbott18-10-bh} and $7.8\pm1.2\msun$ \citep{Ozel2010}, respectively. However core-collapse supernova (SN) explosion models with long explosion timescales and significant fallback presented in \cite{Belczynski2012,Fryer2012} can produce remnants with a continuum mass spectrum. Also a recent measurement of a BH with mass $\sim 3.3^{+2.8}_{-0.7}$ \citep{Thompson2019} and candidates reported by LVC with at least one component having a mass between $3\msun$ and $5\msun$ \citep{190924h,190930s} seem to support the hypothesis of the absence of the “mass gap''.

\textcolor{black}{In Fig. \ref{fig:chirp} we show the $M_1-M_2$ configurations compatible with the chirp mass of GW190425 measured in low-latency \citep{GW190425}.}
The vertical lines in Fig. \ref{fig:chirp} indicate the maximum NS mass for two selected EoS: “APR4'' \citep{Akmal1998,Read2009} and “DD2'' \citep{DD2,DD2_2}. They are, respectively, one of the softest and one of the stiffest among the EoS consistent with the constraints from \textcolor{black}{both GW170817/GW190425 \citep{GW170817_2019,Kiuchi2019,Radice2018_3,GW190425} and NICER \citep{Miller2019,Riley2019} results\footnote{\textcolor{black}{In \cite{Barbieri2019_3} we considered only the SFHo EoS for generating the kilonova light curves.}}}. The APR4 EoS gives $M_\mathrm{NS}^\mathrm{max}=2.08\msun$, while DD2 $M_\mathrm{NS}^\mathrm{max}=2.42\msun$. Configurations on the left of these lines correspond to NSNS binaries, while those on the right are BHNS binaries.

\section{Ejecta from GW190425}
\label{sec:ejecta}

\begin{figure}
    \centering
    \includegraphics{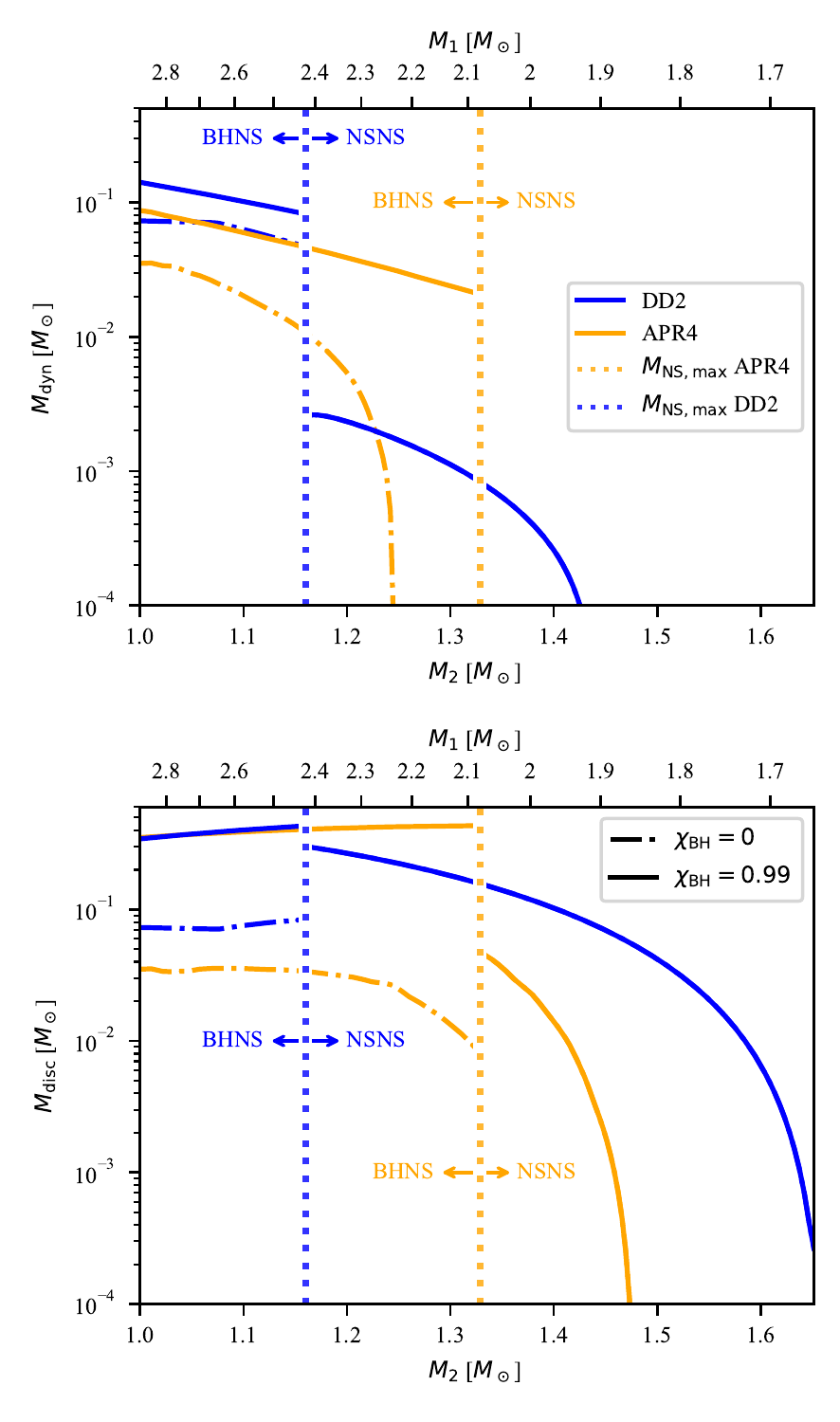}
    \caption{Dynamical ejecta (top) and accretion disc (bottom) mass from binary configurations consistent with the chirp mass of GW190425. Orange and blue lines refer to APR4 and DD2 EoS, respectively. Solid (dot-dashed) line refers to BH spin of 0.99 (0). \textcolor{black}{Dotted vertical lines indicate the maximum NS mass for the two EoS.}
    }
    \label{fig:ejecta}
\end{figure}

In a NSNS merger, partial tidal disruption in the late inspiral phase and crusts impact at the merger produce an outflow of neutron-rich material. Two components can be identified: the dynamical ejecta, which are gravitationally unbound and leave the system, and the accretion disc, the gravitationally bound component around the merger remnant. Additional outflows can arise from the accretion disc: the “wind ejecta'' produced by magnetic pressure and neutrino-matter interaction and the “secular ejecta'' produced by viscous processes \citep[e.g.][]{Dessart2009,Metzger2010,Metzger2014,Perego2014,Siegel2014,Just2015,Siegel2017,Fujibayashi2018}.

\textcolor{black}{The radioactive decay of elements produced in these ejecta through $r$-process nucleosynthesis powers the kilonova (KN hereon) emission \citep{Lattimer1974,Li1998,Metzger2017}. In order to calculate the mass in dynamical ejecta we use the fitting formulae presented in \cite{Radice2018_2} (calibrated on a set of high-resolution general-relativistic hydrodynamic simulations).} 

\textcolor{black}{For symmetric mergers \cite{Radice2018_2} found that $M_\mathrm{disc}$ can be calculated as a function of only the binary dimensionless tidal deformability parameter $\tilde{\Lambda}$ \citep{Raithel2018}. As shown in Fig. \ref{fig:chirp}, we are considering asymmetric NSNS mergers. For these binary configurations, \cite{Kiuchi2019} found that the fitting formula in \cite{Radice2018_2} underestimates the accretion disc masses, indicating that  $M_\mathrm{disc}$ must be calculated as a function of $\tilde{\Lambda}$ and the mass ratio $q$. Thus, to estimate the disc mass we present and adopt a new fitting formula described in Appendix \ref{ap:fit_Om} \citep[for further description and application of this formula see][]{Salafia2020}, based on results from the numerical simulations presented in \cite{Radice2018_2} and \cite{Kiuchi2019}. This new fitting formula gives values in good agreement with simulations of both symmetric and asymmetric mergers. Therefore both $M_\mathrm{dyn}$ and $M_\mathrm{disc}$ depend on the NS masses and tidal deformabilities.}

\textcolor{black}{We neglect the possibility of an energy injection in the ejecta from a remnant NS state. The NSNS systems considered here carry large total masses and likely  collapse promptly to a BH. Alternatively, due to their large masses, an intermediate hyper-massive NS phase will have a short lifetime. We use the recently published fitting formulae by \cite{Bauswein2020} to calculate the threshold mass $M_\mathrm{thr}$ for prompt collapse of the binary into a BH. We find that for APR4 EoS all configurations promptly form a BH, while for DD2 EoS  $\sim 64$\% of the configurations undergo prompt collapse.}

Also BHNS mergers are expected to produce dynamical ejecta and accretion discs, if the NS suffers partial tidal disruption before plunging into the BH \citep{Rosswog2005,Kyutoku2011,Foucart2013}. We calculate the dynamical ejecta and accretion disc properties adopting the fitting formulae from \cite{Kawaguchi2016} and \cite{Foucart2018}. We follow \cite{Barbieri2019} to use as fundamental parameters the BH and NS masses, the BH spin and the NS tidal deformability\footnote{We note that the fitting formula from \cite{Kawaguchi2016} for the mass of dynamical ejecta also depends on $\iota$, which is the angle between the BH spin and the total binary angular momentum. In this work we consider $\iota=0$, corresponding to non-precessing binaries.}.

As discussed in \cite{Barbieri2019}, fixing all the other binary parameters, the larger the BH spin is the more ejecta are produced. Therefore in order to obtain the lower and upper bound on possible ejecta production from GW190425, we assume for the BHNS configurations two spin values: $\chi_\mathrm{BH}=0$ and $\chi_\mathrm{BH}=0.99$. Fig. \ref{fig:ejecta} shows the dynamical ejecta (top) and accretion disc (bottom) masses for configurations consistent with the inferred chirp mass for GW190425. 

BHNS configurations are represented by blue curves on the left of the blue dotted vertical line (DD2 EoS, $M_1>2.42\msun$) and orange curves on the left of the orange dotted vertical line (APR4 EoS, $M_1>2.08\msun$). Different line styles indicate the different BH spin values. It is clear that BHNS mergers characterized by small mass ratios and low-mass (large-$\Lambda_\mathrm{NS}$) NSs represent the optimal combination for ejecta production. Indeed in these cases we expect massive dynamical ejecta and discs for both EoSs and both BH spins. For DD2 EoS, BHNS mergers with $\chi_\mathrm{BH}=0$ ($\chi_\mathrm{BH}=0.99$) produce $M_\mathrm{dyn}\sim6-7\times10^{-2}\msun$ and $M_\mathrm{disc}\sim7-8\times10^{-2}\msun$ ($M_\mathrm{dyn}\sim10^{-1}\msun$ and $M_\mathrm{disc}\sim4\times10^{-1}\msun$). For APR4 EoS, BHNS mergers with $\chi_\mathrm{BH}=0.99$ produce $5\times10^{-2}\lesssim M_\mathrm{dyn}\lesssim9\times 10^{-2}\msun$ and $M_\mathrm{disc}\sim4\times10^{-1}\msun$. Instead for $\chi_\mathrm{BH}=0$ they produce $10^{-2}\msun\lesssim M_\mathrm{disc}\lesssim3\times10^{-2}\msun$, while dynamical ejecta with $10^{-3}\msun\lesssim M_\mathrm{dyn}\lesssim3\times10^{-2}$ are produced only for $M_\mathrm{BH}\gtrsim2.3\msun$.

NSNS configurations are represented by blue curves on the right of the blue dotted vertical line (DD2 EoS, $M_1<2.42\msun$) and orange curves on the right of the orange dotted vertical line (APR4 EoS, $M_1<2.08\msun$). These configurations are the worst for dynamical ejecta production, since massive NSs have small tidal deformability. No dynamical ejecta are produced for APR4 EoS and $M_\mathrm{dyn}<3\times10^{-3}$ for DD2 EoS. For what concerns $M_\mathrm{disc}$, asymmetric NSNS binaries produce discs in between the $\chi_\mathrm{BH}=0$ and $\chi_\mathrm{BH}=0.99$ cases, namely $M_\mathrm{disc}\sim2\times10^{-1}\msun$ for DD2 EoS and $M_\mathrm{disc}\sim4\times10^{-2}\msun$. Moving toward symmetric NSNS binaries ($q\to1$), the disc mass significantly decreases (for APR4 EoS no disc is even produced for $q\lesssim1.27$).

In the following section we show how the differences in the ejecta properties leads to different KNae luminosities for the BHNS and NSNS case.

\section{Kilonova of GW190425}
\label{sec:kilonova}

We compute the KN light curves using the semi-analytical model\footnote{We tested our model against GW170817: multi-wavelength KN light curves obtained with our model using the parameters inferred for this event \citep{GW170817_dynamical,Perego2017} are consistent with the observations \citep{Villar2017}. Moreover, our light curves peak magnitudes and time behaviour are consistent with \cite{Kawaguchi2019}, that derived NSNS/BHNS KN light curves from radiative transfer simulations (including multiple ejecta components effects).} presented in \cite{Barbieri2020} \citep[in part based on ][]{Grossman2014,Martin2015,Perego2017}. This model adopts fitting formulae that provide the mass in the ejecta, presented in \citealt{Kawaguchi2016,Foucart2018} for BHNS and \citealt{Radice2018_2,Salafia2020} for NSNS).
\textcolor{black}{In Table \ref{tab:table_params} we list the assumed model parameters for NSNS mergers \citep[according to][]{Perego2017} and for BHNS mergers \citep[according to][]{Kawaguchi2016,Fernandez2017,Just2015}.}

\begin{table*}
\centering
\begin{tabular}{|*{4}{c|}} 
\hline
\multicolumn{1}{|c}{Parameter} &
\multicolumn{1}{|c|}{Description}&
\multicolumn{1}{|c|}{NSNS}&
\multicolumn{1}{|c|}{BHNS}\\
\cline{1-4}
\multicolumn{1}{|c}{$\xi_\mathrm{w}$} &
\multicolumn{1}{|c|}{Accretion disc mass fraction flowing in wind ejecta} &
\multicolumn{1}{|c|}{0.05}
&
\multicolumn{1}{|c|}{0.01} \\
\hline
\multicolumn{1}{|c}{$\xi_\mathrm{s}$} &
\multicolumn{1}{|c|}{Accretion disc mass fraction flowing in secular ejecta} &
\multicolumn{1}{|c|}{0.2}&
\multicolumn{1}{|c|}{0.2} \\
\hline
\multicolumn{1}{|c}{$v_\mathrm{w}$} &
\multicolumn{1}{|c|}{Wind ejecta velocity} &
\multicolumn{1}{|c|}{0.067 c}&
\multicolumn{1}{|c|}{0.1 c} \\
\hline
\multicolumn{1}{|c}{$v_\mathrm{s}$} &
\multicolumn{1}{|c|}{Secular ejecta velocity} &
\multicolumn{1}{|c|}{0.04 c}&
\multicolumn{1}{|c|}{0.1 c} \\
\hline
\multicolumn{1}{|c}{$k_\mathrm{d}$} &
\multicolumn{1}{|c|}{Dynamical ejecta opacity} &
\multicolumn{1}{|c|}{15 cm$^2$/g}&
\multicolumn{1}{|c|}{15 cm$^2$/g} \\
\hline
\multicolumn{1}{|c}{$k_\mathrm{w}$} &
\multicolumn{1}{|c|}{Wind ejecta opacity} &
\multicolumn{1}{|c|}{0.5 cm$^2$/g}&
\multicolumn{1}{|c|}{1 cm$^2$/g} \\
\hline
\multicolumn{1}{|c}{$k_\mathrm{s}$} &
\multicolumn{1}{|c|}{Secular ejecta opacity} &
\multicolumn{1}{|c|}{5 cm$^2$/g}&
\multicolumn{1}{|c|}{5 cm$^2$/g} \\
\hline
\multicolumn{1}{|c}{$\theta_\mathrm{d}$} &
\multicolumn{1}{|c|}{Dynamical ejecta latitudinal opening angle (from the equatorial plane)} &
\multicolumn{1}{|c|}{80 deg}&
\multicolumn{1}{|c|}{17 deg} \\
\hline
\multicolumn{1}{|c}{$\phi_\mathrm{d}$} &
\multicolumn{1}{|c|}{Dynamical ejecta azimuthal opening angle} &
\multicolumn{1}{|c|}{$2\pi$ rad}&
\multicolumn{1}{|c|}{$\pi$ rad} \\
\hline
\multicolumn{1}{|c}{$\theta_\mathrm{w}$} &
\multicolumn{1}{|c|}{Wind ejecta opening angle (from the polar axis)} &
\multicolumn{1}{|c|}{60 deg}&
\multicolumn{1}{|c|}{60 deg} \\
\hline
\end{tabular}
\caption{\textcolor{black}{Assumed ejecta properties for NSNS and BHNS mergers.}}
\label{tab:table_params}
\end{table*}

\begin{figure*}
    \centering
    \includegraphics{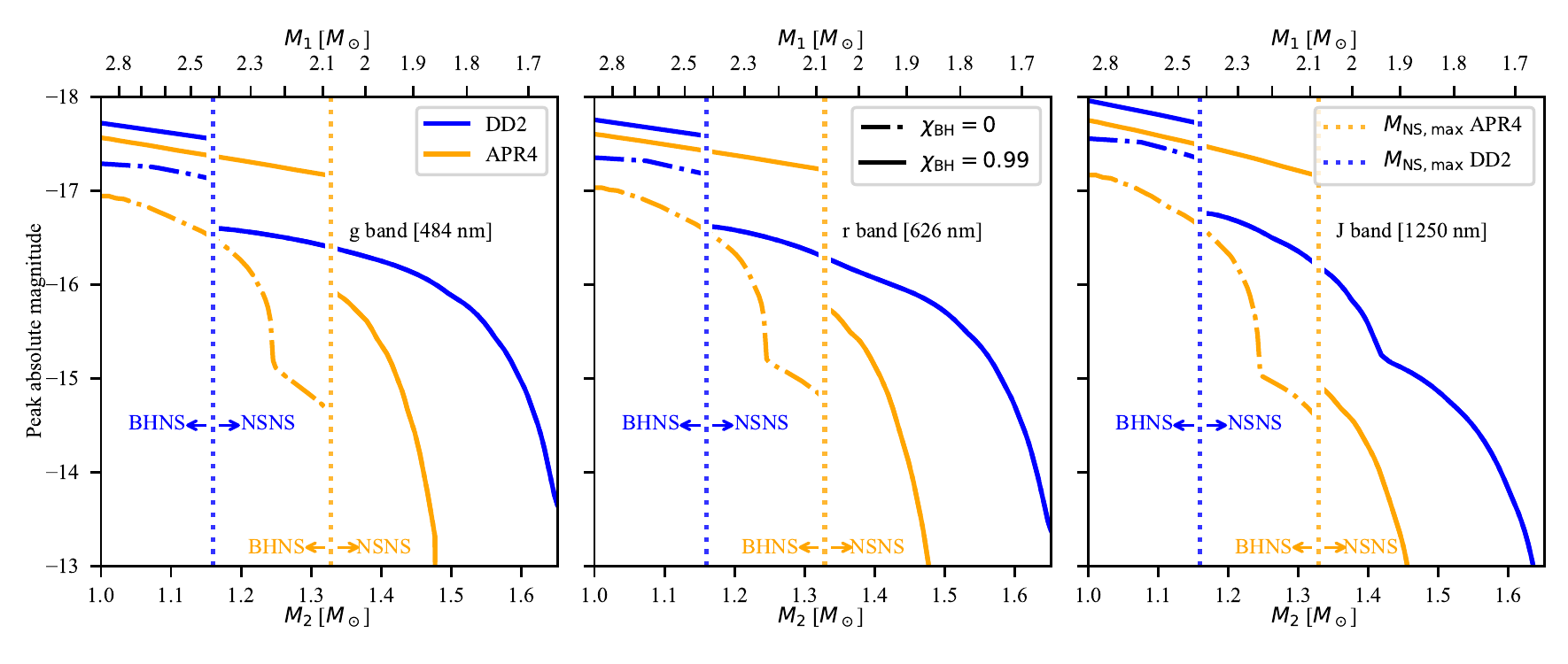}
    \caption{Peak absolute magnitude of KNae from binary configurations consistent with the chirp mass of GW190425. Left, central and right panels refer to, respectively, \textit{g} (484 nm), \textit{r} (626 nm) and \textit{J} (1250 nm) bands. Orange (blue) line refers to APR4 (DD2) EoS. Solid and dot-dashed lines refer to BH spin of 0.99 and 0, respectively. Dotted vertical lines indicate the maximum NS mass for the two EoS.}
    \label{fig:kn}
\end{figure*}

\begin{figure*}
    \centering
    \includegraphics{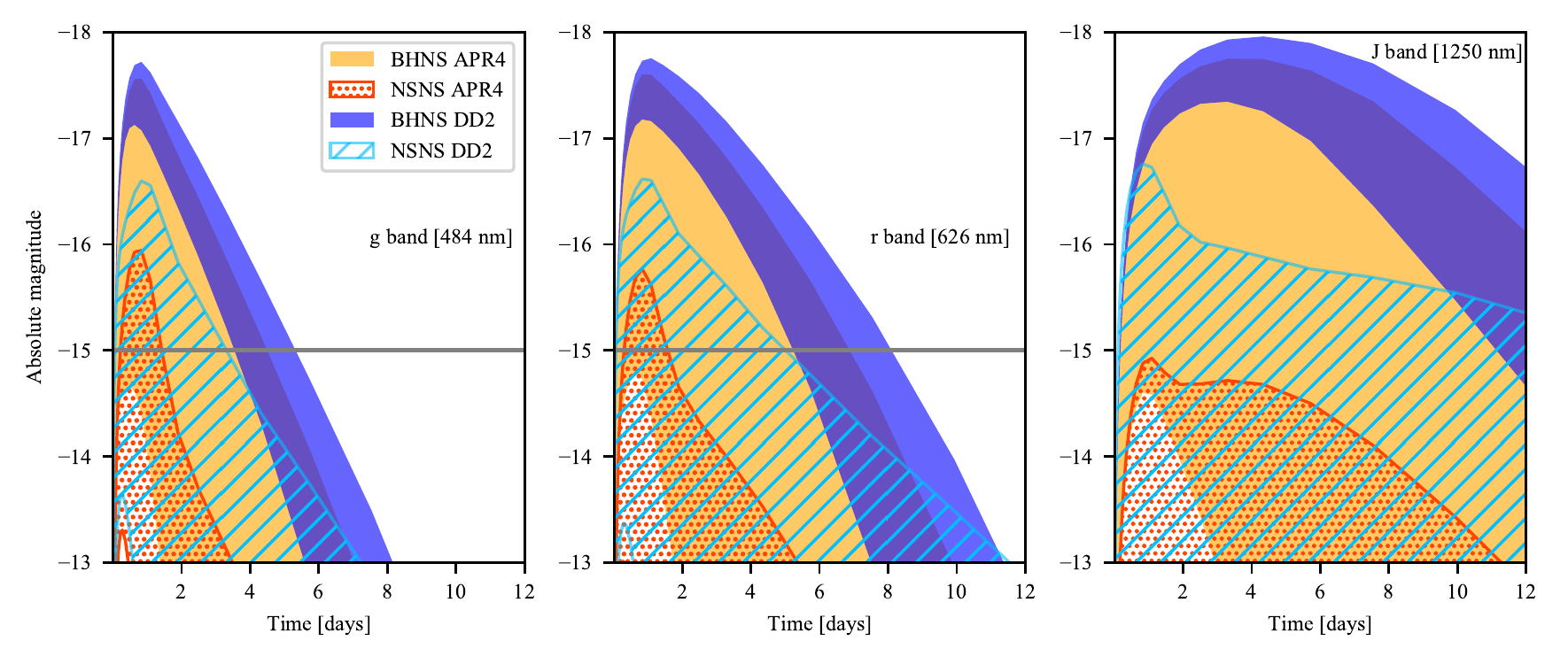}
    \caption{KN light curves ranges for binary configurations consistent with the chirp mass of GW190425. For BHNS cases upper bounds are obtained considering $\chi_\mathrm{BH}=0.99$, while lower bounds are obtained considering $\chi_\mathrm{BH}=0$. Left, central and right panels refer to, respectively, \textit{g} (484 nm), \textit{r} (626 nm) and \textit{J} (1250 nm) bands. Orange (blue) region refers to BHNS mergers for APR4 (DD2) EoS. Dark orange dotted and light blue hatched regions refer to NSNS mergers for APR4 and DD2 EoS, respectively. Gray horizontal lines correspond to the limiting magnitude in GW190425 EM follow-up with ZTF, assuming a distance $d_\mathrm{L}=161$ Mpc.}
    \label{fig:range_kn}
\end{figure*}

In Fig. \ref{fig:kn} we show the peak absolute magnitude of KNae in three relevant bands ({\it g, r, J}) from binary configurations consistent with the  chirp mass inferred for GW190425. Obviously the KN luminosity reflects the ejecta properties (similar trends in this figure and Fig. \ref{fig:ejecta}). We find that there is a difference of $\sim1-1.5$ magnitudes at peak between the most luminous KNae from BHNS and NSNS mergers.

In Fig. \ref{fig:range_kn} we show the KN light curves ranges for the same bands and for binary configurations consistent with GW190425--$M_\mathrm{c}$. For BHNS cases, the lower bounds are obtained considering non-spinning BHs ($\chi_\mathrm{BH}=0$), while the upper bounds are obtained considering maximally-rotating BHs ($\chi_\mathrm{BH}=0.99$). For the DD2 EoS, BHNS KNae are brighter than the NSNS case at early times (from $\sim$ hours to $\sim3$ days for \textit{g} band, $\sim4$ days for \textit{r} band and $\sim1$ week for \textit{J} band). For the APR4 EoS the KN ranges for NSNS mergers overlap with the BHNS ones in the lower (low-luminosity) region. However many BHNS configurations produce brighter KNae with respect to NSNS cases. In particular for the \textit{J} band in the first $\sim20$ hours all the BHNS KNae are brighter than NSNS ones.

In Fig. \ref{fig:range_kn} we also show the limiting magnitude in the GW190425 EM follow-up with the Zwicky Transient Facility \citep[ZTF, ][]{Bellm2019,Coughlin2019_2} in the \textit{g} and \textit{r} bands, assuming that the merger happened at a distance $d_\mathrm{L}=161$ Mpc \citep{GW190425}. We find that BHNS KNae would have been detectable for all (almost all) the binary configurations\footnote{\textcolor{black}{We compared our results with a recent work on the possibility that GW190425 was a BHNS merger \citep[][appeared on arXiv during the writing of this paper]{Kyutoku2020}. Like us, they too find that the KN associated with a BHNS merger consistent with the chirp mass of GW190425 could have been detected during the EM follow-up.}} for DD2 (APR4) EoS in the first $\sim 4-5$ days. Some NSNS configurations for APR4 (DD2) EoS would have produced detectable KN, even if close to the limiting magnitude, for the first $\sim 2$ ($\sim 4-6$) days.

\subsection{\textcolor{black}{Kilonovae from different binary configurations}}
\label{ap:config}
\textcolor{black}{In this section we focus on how the light curves from different binary configurations are distributed in the magnitude-time domain, in order to better understand the ranges' overlaps in Fig. \ref{fig:range_kn}.
Assuming flat distributions for $M_\mathrm{NS}$, $M_\mathrm{BH}$ and $\chi_\mathrm{BH}$, for each EoS we select some NSNS/BHNS configurations coherent with GW190425--$M_\mathrm{c}$ and we show the corresponding KN light curves. The selected configurations are reported in Table \ref{tab:xxx}.
\begin{table}
\centering
\begin{tabular}{|*{4}{c|}} 
\hline
\multicolumn{2}{|c}{} &
\multicolumn{1}{|c|}{$M_1$ [$\msun$]}&
\multicolumn{1}{|c|}{$M_2$ [$\msun$]}\\
\hline
\multicolumn{2}{|c}{} &
\multicolumn{1}{|c|}{1.90} &
\multicolumn{1}{|c|}{1.45} \\
\cline{3-4}
\multicolumn{2}{|c}{NSNS (APR4)} &
\multicolumn{1}{|c|}{1.95} &
\multicolumn{1}{|c|}{1.41} \\
\cline{3-4}
\multicolumn{2}{|c}{} &
\multicolumn{1}{|c|}{2.00} &
\multicolumn{1}{|c|}{1.38} \\
\hline
\multicolumn{2}{|c}{} &
\multicolumn{1}{|c|}{1.70} &
\multicolumn{1}{|c|}{1.61} \\
\cline{3-4}
\multicolumn{2}{|c}{} &
\multicolumn{1}{|c|}{1.80} &
\multicolumn{1}{|c|}{1.52} \\
\cline{3-4}
\multicolumn{2}{|c}{} &
\multicolumn{1}{|c|}{1.90} &
\multicolumn{1}{|c|}{1.45} \\
\cline{3-4}
\multicolumn{2}{|c}{NSNS (DD2)} &
\multicolumn{1}{|c|}{2.00} &
\multicolumn{1}{|c|}{1.38} \\
\cline{3-4}
\multicolumn{2}{|c}{} &
\multicolumn{1}{|c|}{2.10} &
\multicolumn{1}{|c|}{1.32} \\
\cline{3-4}
\multicolumn{2}{|c}{} &
\multicolumn{1}{|c|}{2.20} &
\multicolumn{1}{|c|}{1.26} \\
\cline{3-4}
\multicolumn{2}{|c}{} &
\multicolumn{1}{|c|}{2.30} &
\multicolumn{1}{|c|}{1.21} \\
\cline{3-4}
\multicolumn{2}{|c}{} &
\multicolumn{1}{|c|}{2.40} &
\multicolumn{1}{|c|}{1.17} \\
\hline
\multicolumn{2}{|c}{} &
\multicolumn{1}{|c|}{2.20} &
\multicolumn{1}{|c|}{1.26} \\
\cline{3-4}
\multicolumn{2}{|c}{BHNS (APR4)} &
\multicolumn{1}{|c|}{2.40} &
\multicolumn{1}{|c|}{1.17} \\
\cline{3-4}
\multicolumn{2}{|c}{} &
\multicolumn{1}{|c|}{2.60} &
\multicolumn{1}{|c|}{1.09} \\
\cline{3-4}
\multicolumn{2}{|c}{} &
\multicolumn{1}{|c|}{2.80} &
\multicolumn{1}{|c|}{1.03} \\
\hline
\multicolumn{2}{|c}{} &
\multicolumn{1}{|c|}{2.50} &
\multicolumn{1}{|c|}{1.13} \\
\cline{3-4}
\multicolumn{2}{|c}{BHNS (DD2)} &
\multicolumn{1}{|c|}{2.70} &
\multicolumn{1}{|c|}{1.06} \\
\cline{3-4}
\multicolumn{2}{|c}{} &
\multicolumn{1}{|c|}{2.90} &
\multicolumn{1}{|c|}{1.00} \\
\hline
\end{tabular}
\caption{\textcolor{black}{Selected NSNS/BHNS configurations coherent with GW190425--$M_\mathrm{c}$ for the analysis in \S~\ref{ap:config}.}}
\label{tab:xxx}
\end{table}
Thus we select equally spaced primary masses and we calculate the corresponding $M_2$ using the chirp mass (for NSNS systems with APR4 EoS we start from $M_1=1.9$ $\msun$ because configurations with a less massive primary do not produce any emission). For BHNS configurations, we assume three spin values: $\chi_\mathrm{BH}=0,0.5,0.99$.
\\In the first row of Fig. \ref{fig:config} we show the KN light curves for the selected NSNS configurations and the expected ranges from Fig. \ref{fig:range_kn}. We find that KN emission from the different configurations almost uniformly cover the expected range. In the second row of Fig. \ref{fig:config} we show the KN light curves for BHNS systems assuming APR4 EoS and the corresponding NSNS KN range from Fig. \ref{fig:range_kn}. We find that the majority of BHNS KNae are more concentrated in the bright region, while only few light curves fall in the dim region overlapping with the NSNS KN range (in particular, those corresponding to low spin and small mass ratio). Therefore the overlap at almost all times between BHNS and NSNS KN expected ranges for APR4 EoS presented in Fig. \ref{fig:range_kn} is in reality limited only to few configurations. The same holds for the late time overlaps for DD2 EoS (bottom row of Fig. \ref{fig:config}). This strengthens the possibility to distinguish the nature of the \emph{ambiguous} merging system through the observation of the associated KN}.\footnote{\textcolor{black}{We stress that for simplicity we assumed flat distributions on $M_\mathrm{NS}$, $M_\mathrm{BH}$ and $\chi_\mathrm{BH}$. Hopefully future observations will better constrain the distributions for these parameters, specially for $\chi_\mathrm{BH}$.}} 

\subsection{\textcolor{black}{KN light curves varying model parameters}}
\label{ap:variation}
\textcolor{black}{In this section we perform the analysis presented in \S~\ref{sec:kilonova} considering some variations in the model parameters. The aim of this section is to test the robustness of our results and their sensitivity to modeling assumptions. We consider three variations:
\begin{itemize}
    \item \textit{Variation 1 (V1):} as explained above, the NSNS configurations corresponding to GW190425-$M_\mathrm{c}$ involve mostly asymmetric binaries. \cite{Bernuzzi2020} recently found that asymmetric NSNS mergers produce dynamical ejecta with a crescent-like geometry, similarly to BHNS mergers \citep{Kawaguchi2016}. Thus in V1 we set the NSNS dynamical ejecta geometrical parameters to $\theta_\mathrm{d}=20$ deg and $\phi_\mathrm{d}=\pi$ rad.
    \item \textit{Variation 2 (V2):} besides being mostly asymmetric, NSNS configurations corresponding to GW190425--$M_\mathrm{c}$ also involve massive stars. As explained in \S~\ref{sec:ejecta}, in almost all the cases the merger results in a prompt BH formation, without an intermediate hyper-massive NS phase. The consequent lack of neutrino winds (and neutrino-matter interaction) could lead to less massive wind ejecta with a smaller electron fraction (larger opacity). In such a scenario the ejecta properties would be similar to the BHNS case. Thus in V2 we set the NSNS parameters $\xi_\mathrm{w}$, $v_\mathrm{w}$, $v_\mathrm{s}$, $k_\mathrm{w}$, $\theta_\mathrm{d}$ and $\phi_\mathrm{d}$ to the same values of the BHNS case.
    \item \textit{Variation 3 (V3):} we explore the case in which BHNS mergers produce ejecta with much larger opacities (lower electron fractions). We set $k_\mathrm{d}=30$ cm$^2$/g, $k_\mathrm{w}=5$ cm$^2$/g and $k_\mathrm{s}=15$ cm$^2$/g.  
\end{itemize}
Fig. \ref{fig:variation} shows the analogous of Fig. \ref{fig:range_kn} for V1 (top row), V2 (central row) and V3 (bottom row). For what concerns V1, we find that the different dynamical ejecta geometry does not affect the expected NSNS KN ranges for APR4 EoS and only slightly changes the ranges for DD2 EoS. Indeed for the considered binary configurations, as shown in Fig.~\ref{fig:ejecta}, for the APR4 EoS no dynamical ejecta are produced, while for DD2 EoS they have small masses and the dominant components are the ejecta from the disc. For what concerns V2, we find that the reduced wind ejecta mass and the increase in wind and secular ejecta velocities produce faster-evolving NSNS light curves, as can be seen in the rapid decline of NSNS ranges after the peak. For what concerns V3, we find that increasing the BHNS ejecta opacities produce dimmer light curves (the ranges are shifted of $\sim0.5$ mag).
\\Therefore we find that also for these three different model parameters variations our results remain valid. Indeed for V1 the ranges are very similar to Fig.~\ref{fig:range_kn}, for V2 the overlap between BHNS and NSNS expected ranges is even smaller, while for V3 the overlap is a bit larger. However in each case it remains possible to distinguish the nature of the merging \emph{ambiguous} system through the observation of the KN produced by the merger. This demonstrates the robustness of our results with respect to different assumptions on model parameters.
}

\subsection{\textcolor{black}{Possible ejecta and kilonova of GW190425 with BH masses comparable to NS ones}}
\label{ap:ej_kn_bh_mass_ns}
\textcolor{black}{In this section we repeat the analysis performed in \S~\ref{sec:ejecta} - \S~\ref{sec:chirp}, allowing the BHs to have masses comparable with those of NSs. Therefore here we adopt a more agnostic approach, describing BHNS binaries without imposing the condition $M_\mathrm{BH}^\mathrm{min}=M_\mathrm{NS}^\mathrm{max}$ but simply considering that the primary component is a BH and the secondary is a NS. Some studies have demonstrated that such a BHNS system would be compatible with GW170817 multi-messenger observations \citep{Hinderer2019,Foucart2019,Coughlin2019}, although the NSNS nature seems more likely. In this case the \emph{ambiguous} chirp masses are represented by all the values smaller than the maximum $M_\mathrm{c}$ for a NSNS system ($\sim1.81$ $\msun$ for APR4 EoS and $\sim2.11$ $\msun$ for DD2). 
\\
Fig. \ref{fig:ejecta2} is the analogous of Fig. \ref{fig:ejecta}, showing the dynamical ejecta (top) and accretion disc (bottom) masses for configurations consistent with the chirp mass of GW190425. For a given $\chi_\mathrm{BH}$, the general trend for the dynamical ejecta is that $M_\mathrm{dyn}$ decreases for more symmetric BHNS configurations. Also $M_\mathrm{disc}$ decreases for $q\to1$, except for systems with large $\chi_\mathrm{BH}$, that always produce massive accretion discs. Obviously the results for NSNS systems and BHNS configurations with $M_\mathrm{BH}\geq M_\mathrm{NS}^\mathrm{max}$ are the same as above. The crucial difference is that now there are BHNS configurations (with low $\chi_\mathrm{BH}$) producing low-massive ejecta or no ejecta at all. This results in a widening of the expected BHNS KN range in the low-luminosity region, as shown in Fig. \ref{fig:range_kn2}. Therefore we find that KN light curves from BHNS mergers with low-spin and very low-mass BHs (below $M_\mathrm{NS}^\mathrm{max}$) can not be distinguished from NSNS case. However the same arguments presented above are still valid also in this case: the detection of a bright KN would be consistent only with BHNS merger and the two candidates ZTF19aarykkb and ZTF19aasck are inconsistent with the expected KN range (see Fig. \ref{fig:data2}).}

\subsection{\textcolor{black}{Possible kilonova of GW190425 from GW signal analysis posterior samples}}
\label{ap:kn_ps}
\textcolor{black}{
In this section we analyse the KN light curves obtained from the posterior samples from GW190425 GW signal analysis (available at \url{https://dcc.ligo.org/LIGO-P2000026/public}). In particular, we consider the samples obtained using the “\texttt{PhenomDNRT}'' waveform approximant and the high-spin prior. In the top row of Fig. \ref{fig:kn_PS} we show with gray lines the KN light curves for samples representing BHNS configurations, assuming that $M_\mathrm{BH}^\mathrm{min}=M_\mathrm{NS}^\mathrm{max}$. In the central row of Fig. \ref{fig:kn_PS} we show with gray lines the KN light curves for all samples considering that the primary object is a BH. In the bottom row of Fig. \ref{fig:kn_PS} we show with gray lines the KN light curves for samples representing NSNS configurations. In each row, blue (orange) lines represent the selected samples consistent with DD2 (APR4) EoS. In the top row of Fig. \ref{fig:kn_PS} we find that, assuming $M_\mathrm{BH}^\mathrm{min}=M_\mathrm{NS}^\mathrm{max}$, the KNae from BHNS systems for different EoS do not overlap with those from NSNS systems. Instead, in the central row of Fig. \ref{fig:kn_PS} we find an overlap in the low-luminosity region. This is due to the presence of binary configurations producing low-mass ejecta for both EoS (as explained in \S~\ref{ap:ej_kn_bh_mass_ns}). The dashed lines in the first two rows of Fig. \ref{fig:kn_PS} represent the KN ranges for all NSNS configurations (black), those consistent with DD2 EoS (aqua) and those consistent with APR4 EoS (orange). Assuming $M_\mathrm{BH}^\mathrm{min}=M_\mathrm{NS}^\mathrm{max}$, we find that for DD2 EoS BHNS KNae are brighter than NSNS ones (at all times in the $g$ and $r$-band, after $t\sim1.5$ days in the $J$-band). For APR4 EoS the BHNS KNae are brighter than NSNS ones for $t\lesssim1$ day in the $g$-band, $t\lesssim3$ days in the $r$-band and $2\lesssim t\lesssim7$ days in the $J$-band, while in the other time intervals there are some overlaps. Therefore we find that some degeneracies are still present also considering the high-latency parameter estimation analysis, although reduced with respect to the low-latency estimates. However the results of the previous analysis are confirmed, as bright KNae would be consistent only with a BHNS merger. If instead we assume that BHs can have masses below $M_\mathrm{NS}^\mathrm{max}$, BHNS and NSNS KN ranges overlap because, as already explained, moderate spin - almost equal mass BHNS binary configurations produce low-mass ejecta. We stress that the KN ranges in Fig. \ref{fig:kn_PS} are slightly different from those in Fig. \ref{fig:range_kn}. This is due to the fact that, in order to select samples consistent with an EoS, we request that $\tilde{\Lambda}^\mathrm{s}-0.05\tilde{\Lambda}^\mathrm{s}\leq\tilde{\Lambda}(M_1^\mathrm{s},M_2^\mathrm{s},\mathrm{EoS})\leq\tilde{\Lambda}^\mathrm{s}+0.05\tilde{\Lambda}^\mathrm{s}$, where $\tilde{\Lambda}^\mathrm{s}$, $M_1^\mathrm{s}$ and $M_2^\mathrm{s}$ are the binary tidal deformability, primary and secondary mass from the samples, respectively.}

\section{EM follow-up strategy with the knowledge of the chirp mass}
\label{sec:chirp}

\begin{figure*}
    \centering
    \includegraphics{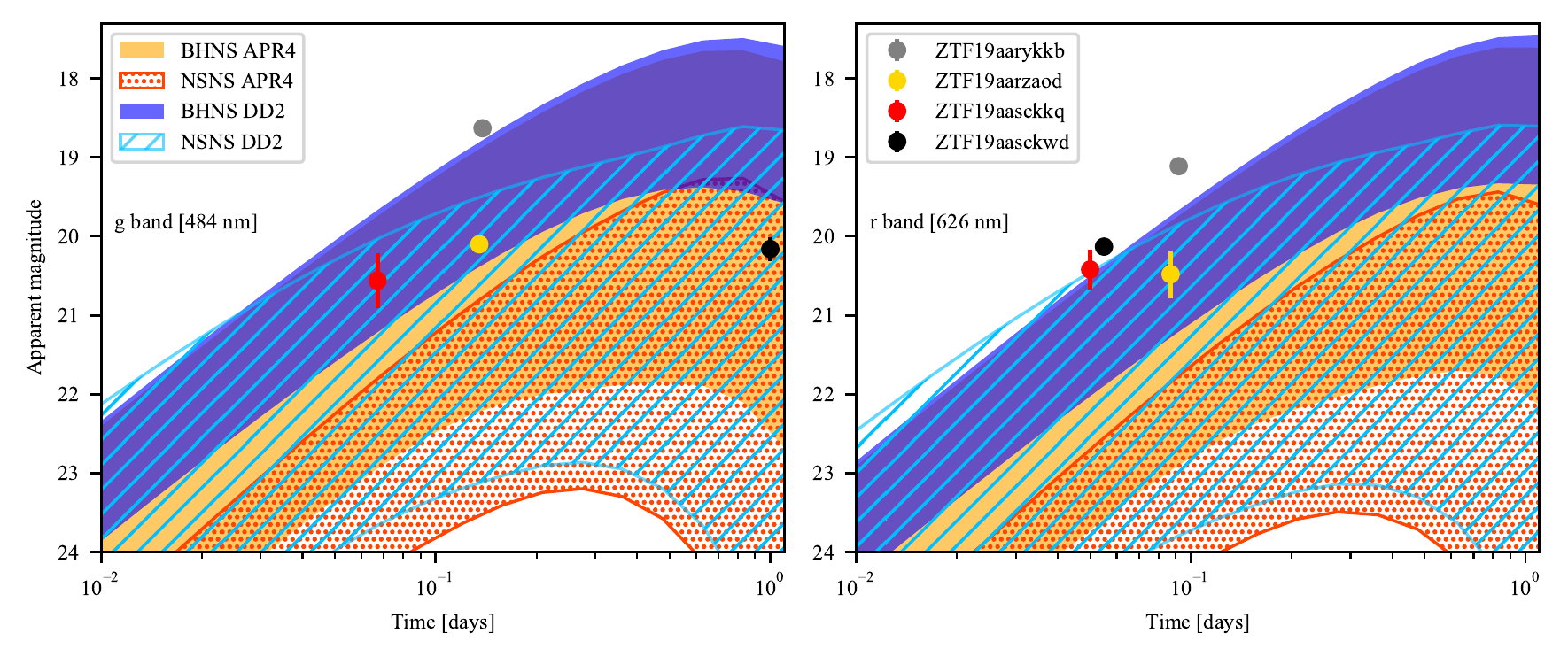}
    \caption{KN light curves ranges for binary configurations consistent with the chirp mass of GW190425. Upper bounds are obtained considering $d_\mathrm{L}=110$ Mpc (and $\chi_\mathrm{BH}=0.99$ for BHNS cases), while lower bounds are obtained considering $d_\mathrm{L}=200$ Mpc (and $\chi_\mathrm{BH}=0$ for BHNS cases). Colored points with errorbars are the first detections by ZTF of promising candidates as EM counterparts to the event. Left and right panels refer to, respectively, \textit{g} (484 nm) and \textit{r} (626 nm) bands. Orange (blue) region refers to BHNS mergers for APR4 (DD2) EoS. Dark orange dotted (light blue hatched) region refers to NSNS mergers for APR4 (DD2) EoS.}
    \label{fig:data}
\end{figure*}

The possibility to distinguish the nature of the merging system for an \emph{ambiguous} event is related to the detection of the associated KN. This is not a simple achievement, as from the analysis of GW signal the uncertainties on the localisation volume (obtained by combining the sky localisation and the distance estimates) can be very large. Thousands of galaxies (and many more transients) could be present in this volume making the identification of the KN associated with the merger very challenging. In the best scenario the KN is identified after some time and the short living/rapidly decaying transients are lost. In the worst scenario the KN is never identified and all the EM counterparts are lost.

In Fig. \ref{fig:range_kn} we show that, knowing the chirp mass, we can calculate the expected KN light curves ranges. This could provide useful criteria to optimize the EM follow-up strategy. Indeed the observation of transients consistent
with KN emission at their first detection could be prioritized for the subsequent photometric and/or spectroscopic follow-up, aimed at classifying them. This could enhance the probability of discovering the electromagnetic counterpart to the GW event.

GW190425 was a single interferometer detection. This is one of the reasons why the sky localisation was poorly informative, being the 90\% credible sky area $\sim 8300$ deg$^2$ \citep{GW190425}\footnote{As a comparison, the GW170817 90\% credible sky area was $\sim 28$ deg$^2$.}. Nonetheless, it is remarkable that the Global Relay of Observatories Watching Transients Happen network observed $\sim21\%$ of the skymap \citep{Coughlin2019_2}. Among all the transients detected during the first 48 hours, 15 candidates were particularly interesting \citep{Kasliwal2019,Anand2019}. After being observed for $\sim$ days they were classified as supernovae (SNe) \citep{Coughlin2019_2}.

In Fig. \ref{fig:data} we show how our argument could be applied to the GW190425 EM follow-up campaign. We calculate the expected apparent magnitude range of KN light curves using the knowledge of the chirp mass $M_\mathrm{c}=1.44\pm0.02\msun$ and the luminosity distance estimate initially circulated by LVC \citep{GCN_GW190425} $d_\mathrm{L}=155\pm45$ Mpc. Considering APR4 or DD2 EoS to describe NS matter, for each of them the lower bound is calculated assuming $\chi_\mathrm{BH}=0$ and $d_\mathrm{L}=200$ Mpc, while the upper bound assuming $\chi_\mathrm{BH}=0.99$ and $d_\mathrm{L}=110$ Mpc. In Fig. \ref{fig:data} we also show the first detections of 4 promising candidates identified by ZTF. These transients were observed for $1-4$ days \citep[see Fig. 3 in ][]{Coughlin2019_2} before being classified as SNe. The first detection of the transients ZTF19aarzaod and ZTF19aasckkq is consistent with the expected KN ranges, thus subsequent observations would have been anyway needed to understand their nature. Instead the transients ZTF19aarykkb and ZTFaasckwd are inconsistent with the expected KN ranges. Therefore other candidates (consistent with the expected range at the moment of their first detection) could have been observed with higher priority.

We are quite confident in defining ZTF19aarykkb and ZTFaasckwd as inconsistent to be the GW190425 counterpart. Indeed these transients would be brighter than the KN produced by a merger whose chirp mass is the one inferred for GW190425, that happened at the lower bound of the luminosity distance $1\sigma$ interval, where the BH is maximally rotating and the NS EoS is one of the stiffest (DD2) among those consistent with GW170817 event.

\begin{figure*}
    \centering
    \includegraphics{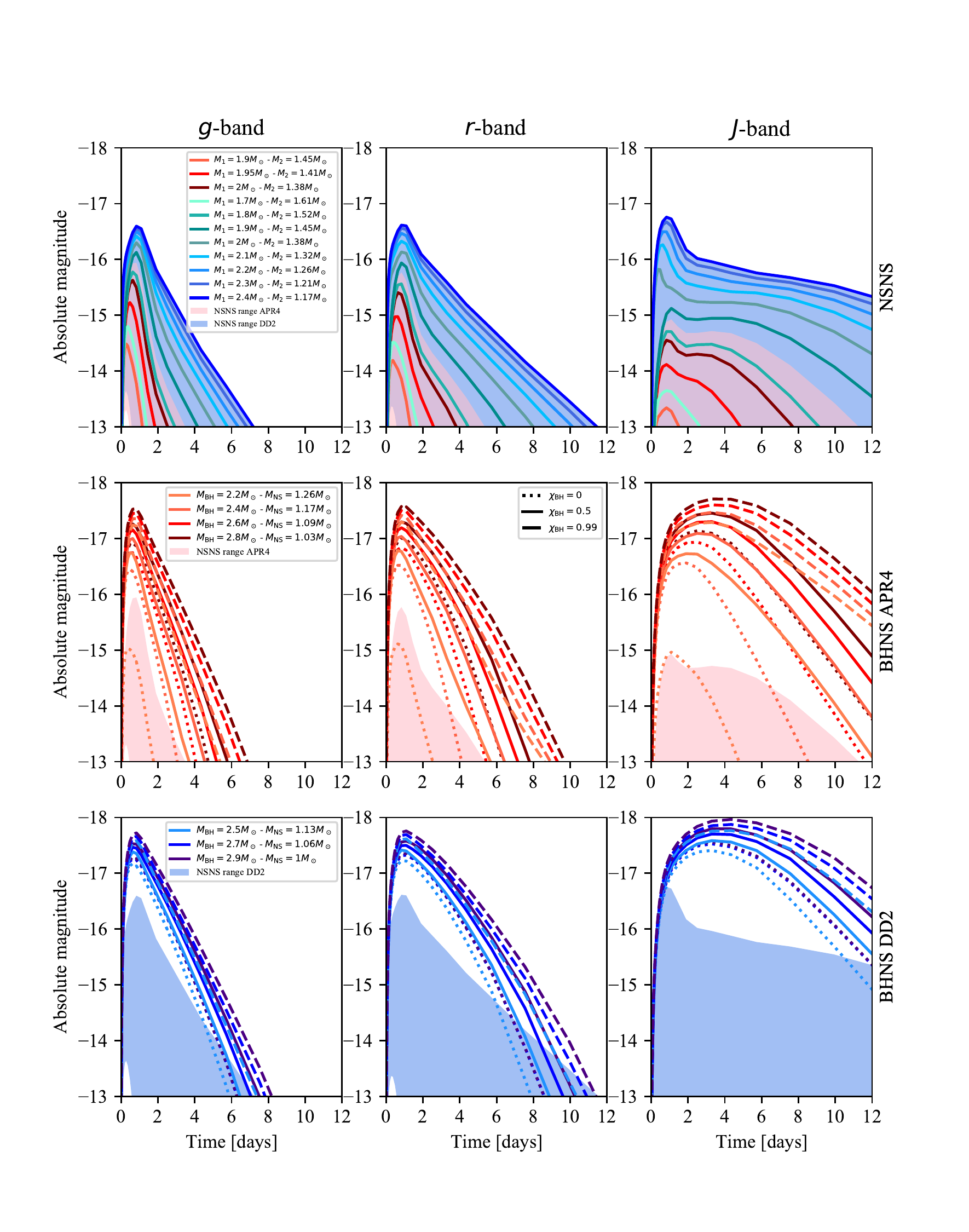}
    \caption{\textcolor{black}{\textit{Top row:} KN light curves from selected NSNS configurations and expected NSNS KN ranges for APR4 (red tones) and DD2 (blue tones) EoS. \textit{Central (bottom) row:} KN light curves from selected BHNS configurations and expected NSNS KN ranges for APR4 (DD2) EoS. In each panel colors indicate different binary component masses (legend in the first column). Linestyles indicate different BH spins (legend in the central panel).}}
    \label{fig:config}
\end{figure*}

\begin{figure*}
    \centering
    \includegraphics{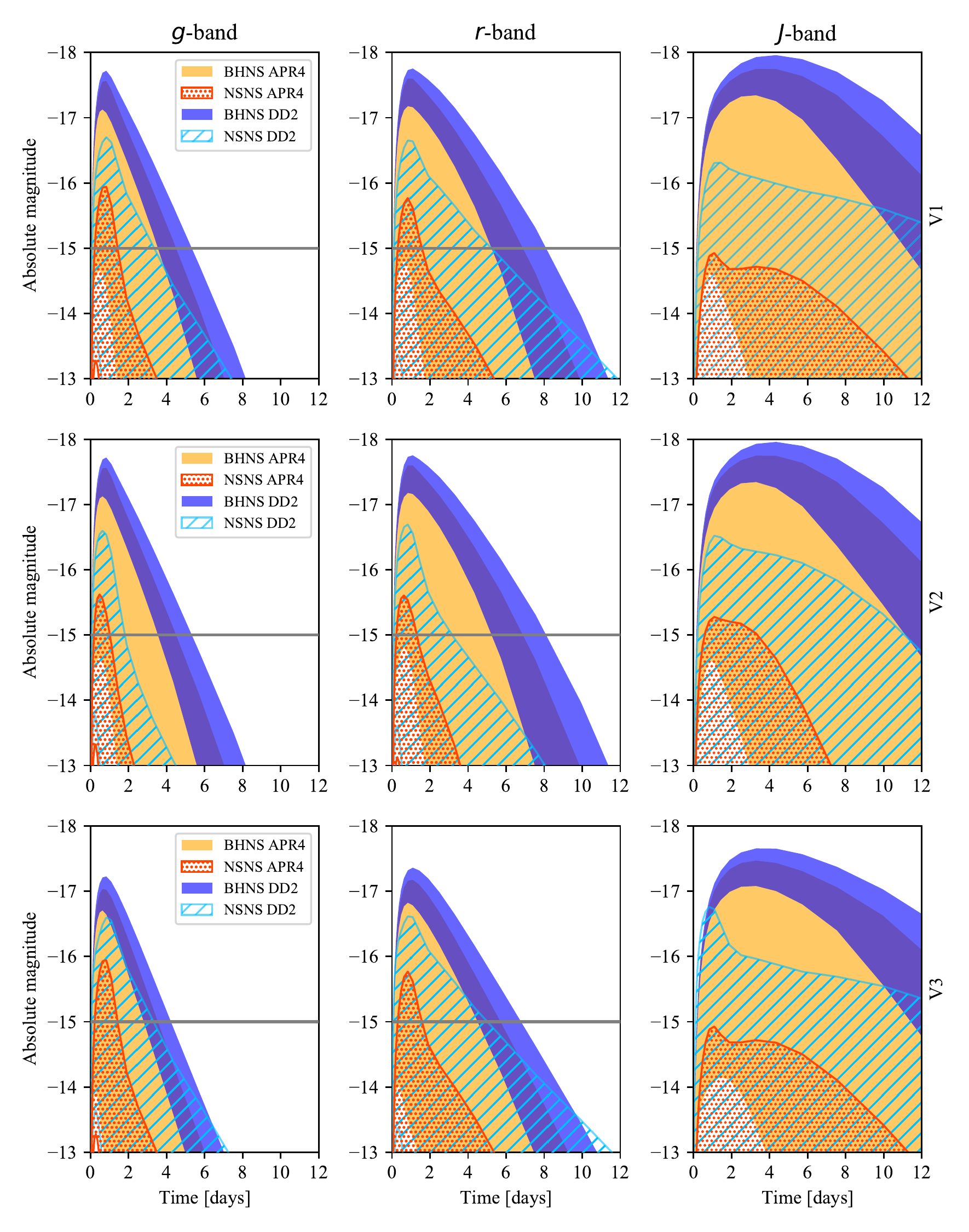}
    \caption{\textcolor{black}{Same as Fig. \ref{fig:range_kn} for model parameters variation V1 (top row), V2 (central row) and V3 (bottom row), respectively. See \S~\ref{ap:variation} for a description of these variations.}}
    \label{fig:variation}
\end{figure*}

\begin{figure*}
    \centering
    \includegraphics[scale=0.95]{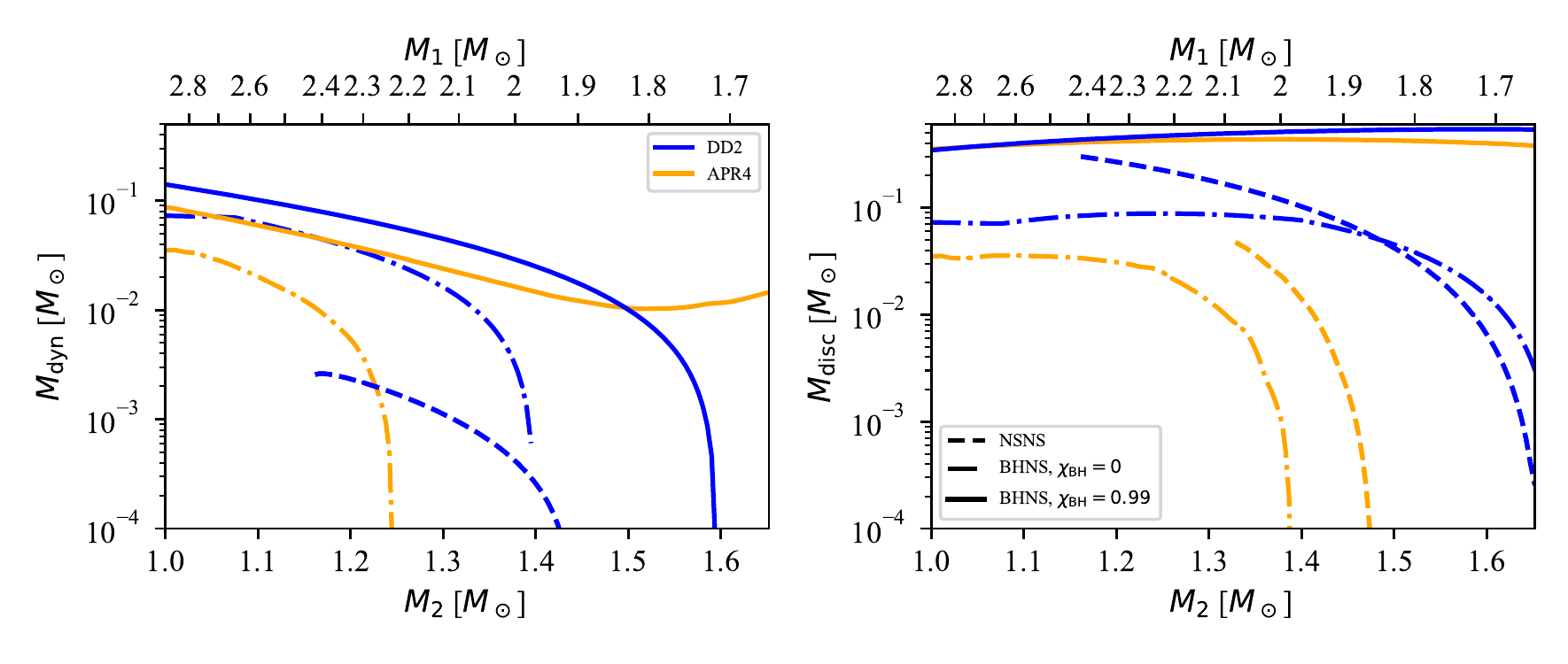}
    \caption{\textcolor{black}{Same as Fig. \ref{fig:ejecta}, assuming BHs with mass below $M_\mathrm{NS}^\mathrm{max}$. Solid (dot-dashed) line refers to BHNS systems with $\chi_\mathrm{BH}=0.99$ (0), while dashed line refers to NSNS.}}
    \label{fig:ejecta2}
\end{figure*}
\begin{figure*}
    \centering
    \includegraphics[scale=0.95]{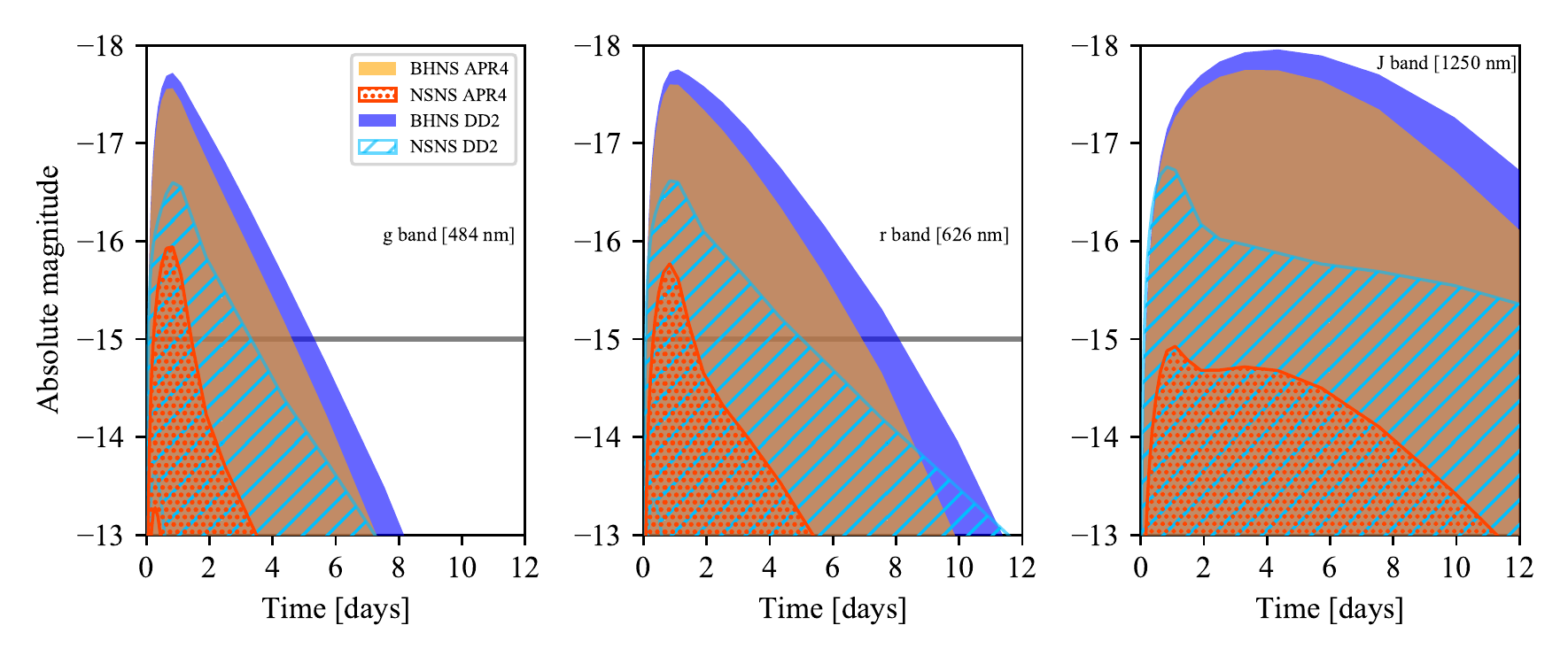}
    \caption{\textcolor{black}{Same as Fig. \ref{fig:range_kn}, assuming BHs with mass below $M_\mathrm{NS}^\mathrm{max}$.}}
    \label{fig:range_kn2}
\end{figure*}
\begin{figure*}
    \centering
    \includegraphics[scale=0.95]{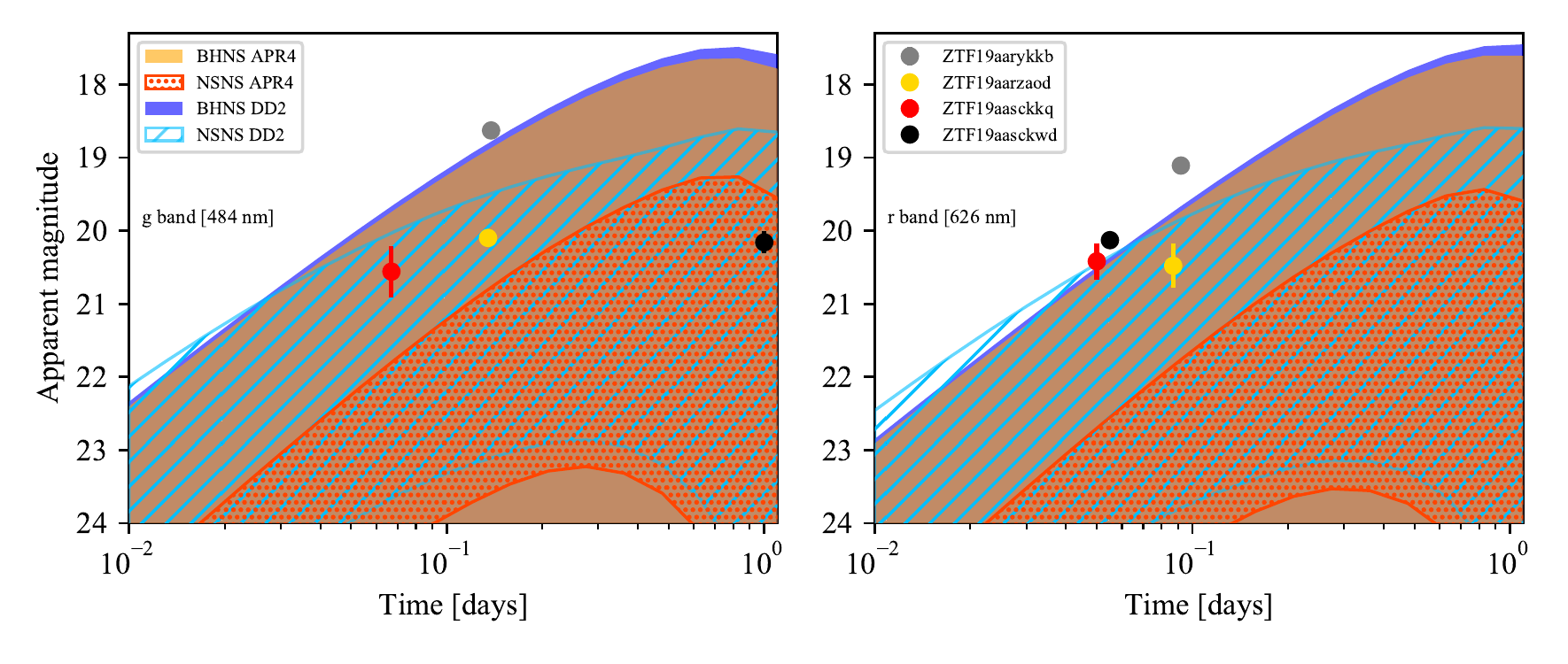}
    \caption{\textcolor{black}{Same as Fig. \ref{fig:data}, assuming BHs with mass below $M_\mathrm{NS}^\mathrm{max}$.}}
    \label{fig:data2}
\end{figure*}

\begin{figure*}
    \centering
    \includegraphics{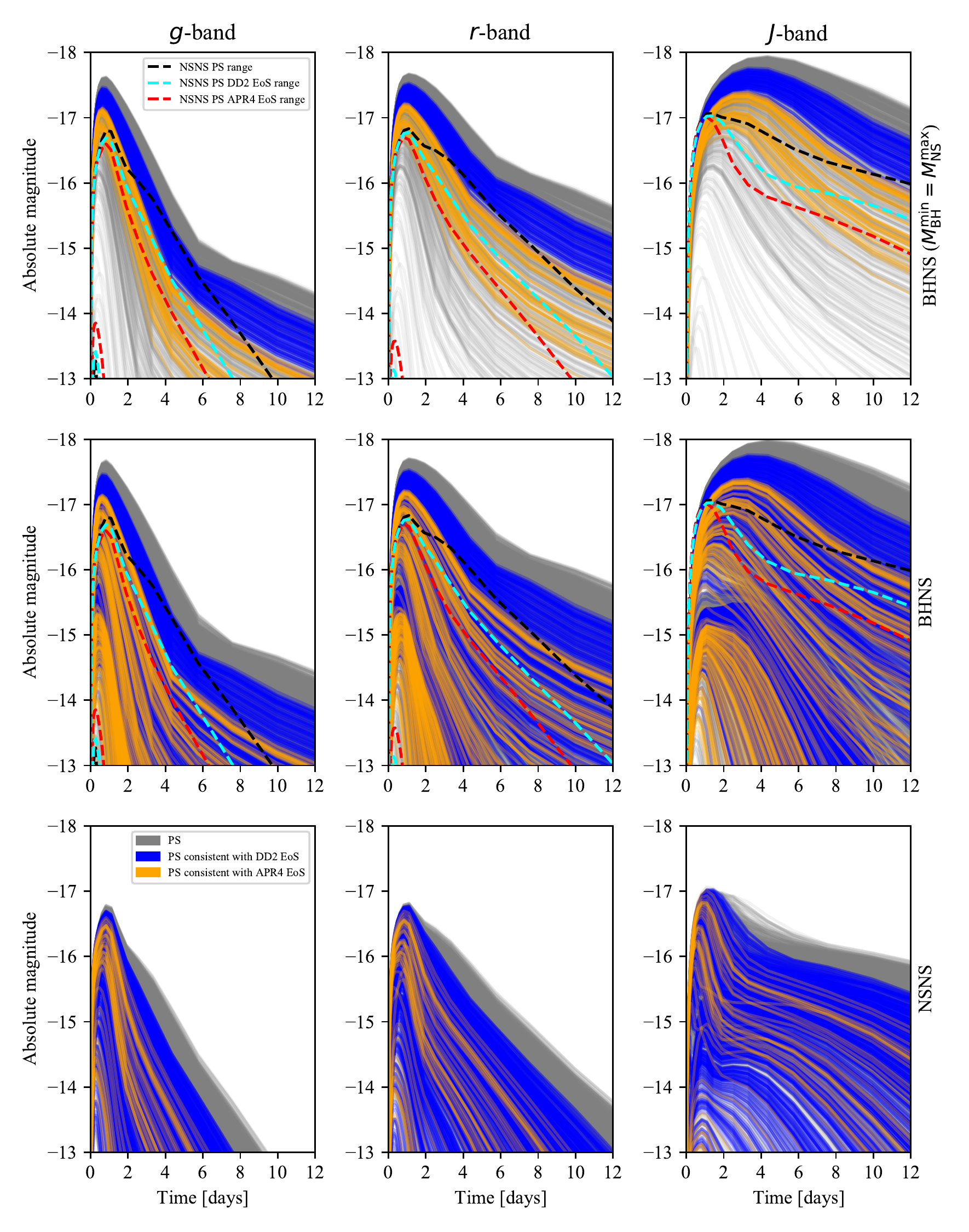}
    \caption{\textcolor{black}{KN light curves for GW190425 posterior samples. \textit{Top row}: samples consistent with BHNS mergers (assuming $M_\mathrm{BH}^\mathrm{min}=M_\mathrm{NS}^\mathrm{max}$). \textit{Central row}: all samples, considering that the primary object is a BH. \textit{Bottom row}: samples consistent with NSNS mergers. Blue (orange) lines indicate samples consistent with DD2 (APR4) EoS. Black, aqua and red dashed lines in the first two rows indicate the NSNS KN ranges for, respectively, all EoS, DD2 and APR4. We consider the $g$-band (left column), $r$-band (central column) and $J$-band (right column).}}
    \label{fig:kn_PS}
\end{figure*}

\section{\textcolor{black}{Summary and results}}
\textcolor{black}{In this work we carried on a low-latency analysis based only on the estimates of the system's chirp mass and luminosity  distance (available few minutes after the trigger). Such analysis helps 
the planning of EM multi-frequency follow-up campaigns, prioritizing the observation of transients to enhance the probability of detecting the EM counterpart. We applied this method to the GW190425 case, constructing NSNS/BHNS kilonova light curve models for that event, considering two equations of state consistent with current constraints from the signals of GW170817/GW190425 and the NICER results, including black hole spin effects and assuming a new formula for the mass of the ejecta. We found  that if our method had been applied to low-latency follow-up of GW190425, two transients (that were observed for $\sim$24 hours before being discarded) would have been immediately discarded (see \S~\ref{sec:chirp}).}

\textcolor{black}{In \S~\ref{sec:kilonova} we showed that if one component of  GW190425 were a BH, the merger could have produced a kilonova far more luminous compared to the NSNS case (examples of kilonova light curves from BHNS mergers as bright as or brighter than NSNS mergers have already been proposed in i.e. \cite{Kawaguchi2019,Barbieri2020}).
 We  further found  that  kilonova  light  curves  from  different NSNS configurations are almost uniformly distributed in the magnitude-time domain, while those from different BHNS configurations  are  more  concentrated  in  the  bright  region. Thus, the overlap presented in Fig. 4 is limited to few configurations strengthening our result. Therefore, the putative observation of kilonova emission associated with GW190425 could have unveiled the nature of the companion to the NS \citep[as suggested in ][]{Barbieri2019_3}.} 

\textcolor{black}{In \S~\ref{ap:config}, \S~\ref{ap:variation} and \S~\ref{ap:ej_kn_bh_mass_ns} we tested the robustness of our results against our model assumptions. Concerning degeneracy, we repeated our analysis on the posterior samples of GW190425 from the high-latency parameter estimation, finding that degeneracy between NSNS and BHNS kilonova light curves is still present, but reduced.  Interestingly, the capability to distinguish the nature of the system using low-latency analysis is comparable to the high-latency case. In \S~\ref{ap:ej_kn_bh_mass_ns} we further found that, if BHs with mass below the maximum mass of NSs exist, the kilonovae from such “very-light'' BHNS systems can be distinguished from the NSNS case only if the BH spin is large.}

\textcolor{black}{Finally, we remark that the identification of a BHNS  merger with \emph{ambiguous} chirp mass would provide the first hint of the existence of “light'' BHs, confuting the presence of a “mass gap'' between NS and BH mass distributions. Such a discovery would have important impact on SN explosion models, favoring those producing a continuum remnant mass spectrum. It would also be crucial for constraining the maximum mass of non-rotating neutron stars.}

\begin{acknowledgements}
We thank F.~Zappa and S.~Bernuzzi for sharing EoS tables. The authors acknowledge support from INFN, under the Virgo-Prometeo initiative. O.~S. acknowledges the Italian Ministry for University and Research (MIUR) for funding through project grant 1.05.06.13. M.~C. acknowledges the COST Action CA16104 “GWverse'', supported by COST (European Cooperation in Science and Technology). During drafting of this paper, M.~C. acknowledges kind hospitality by the Kavli Institute for Theoretical Physics at Santa Barbara, under the program "The New Era of Gravitational-Wave Physics and Astrophysics".
\end{acknowledgements}

\footnotesize{
\bibliographystyle{aa}
\bibliography{references}
}

\appendix

\section{\textcolor{black}{Disc mass fitting formula for NSNS mergers}}
\label{ap:fit_Om}

\subsection{\textcolor{black}{A toy model of mass shedding in asymmetric BNS mergers}}

\begin{figure}
\centering
 \includegraphics[width=0.7\columnwidth]{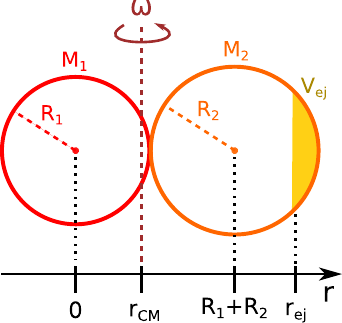}
 \caption{\textcolor{black}{Sketch of the reference geometry in the toy model on which the disc mass fitting formula is based.}}
 \label{fig:toy_model}
\end{figure}

\textcolor{black}{Let us consider a neutron star binary of masses $M_1$ and $M_2$ and radii $R_1$ and $R_2$, right at the moment when the two surfaces touch each other (refer to Fig.~\ref{fig:toy_model} for a sketch of the geometry). Let us neglect the tidal deformation of the two stars for the moment, and let us neglect relativistic effects. Assuming Keplerian orbits, the angular frequency of the binary is $\omega = \sqrt{GM/(R_1+R_2)^3}$, where $M=M_1+M_2$ is the total mass. If we set the origin of our coordinate system at the center of $M_1$, then the center of mass of the system is located at a distance $r_\mathrm{CM}=(R_1+R_2)/(1+M_2/M_1)$ along the line that connects the centers of the two stars. At any point $r>r_\mathrm{CM}$ along this line, the centrifugal acceleration experienced by a co-rotating test mass is 
\begin{equation}
a_\mathrm{c}=\frac{GM}{(R_1+R_2)^2}\frac{r/r_\mathrm{CM}-1}{1+M_1/M_2}
\end{equation}
Now, the ansatz of our toy model is that whenever this centrifugal acceleration exceeds the gravitational acceleration
\begin{equation}
 a_\mathrm{g}=\frac{GM}{(r-r_\mathrm{CM})^2}=\frac{GM}{(R_1+R_2)^2}\frac{(1+M_1/M_2)^2}{(r/r_\mathrm{CM}-1)^2}
\end{equation}
that the merger remnant (assuming no mass loss) would exert at the same distance, then the corresponding part of the star $M_2$ is ejected before the merger. If tidal forces cause the star $M_2$ to stretch to an ellipsoid whose semi-major axis is $\lambda_2 R_2$, then the effect is roughly that of reducing $a_\mathrm{g}$ by $\lambda_2^2$ and increasing $a_\mathrm{c}$ by $\lambda_2$ at the corresponding position. By the condition $a_\mathrm{g}/\lambda_2^2<a_\mathrm{c}\lambda_2$, one obtains that matter beyond $r_\mathrm{ej,2}=(R_1+R_2)[(1+M_1/M_2)^{-1}+\lambda_2^{-1}]$ is ejected. The mass of this matter can be estimated by assuming the neutron star density profile to be uniform, and approximating the volume $V_\mathrm{ej}$ of the ejected matter as a spherical cap (which is reasonable as long as it is small compared to the sphere), which yields
\begin{equation}
 M_\mathrm{d,2}=\frac{V_\mathrm{ej}}{V}M_2\sim \frac{1}{4}(2+x_2)(x_2-1)^2M_2
 \label{eq:Mdisk2}
\end{equation}
where $x_2=(r_\mathrm{ej,2}-R_1-R_2)/R_2$ and we are neglecting the difference between baryon and gravitational mass. If both components have masses not too close to the maximum TOV mass, this can be simplified further by neglecting the difference in neutron star radii. With this assumption,
\begin{equation}
 x_2\sim 2((1+M_1/M_2)^{-1}+\lambda_2^{-1}-1)
 \label{eq:x_2}
\end{equation}
Exchanging $1$ and $2$, one gets the corresponding formula for the disc mass contribution $M_\mathrm{d,1}$ from the star $M_1$, so that the disc mass is eventually $M_\mathrm{d}=M_\mathrm{d,1}+M_\mathrm{d,2}$.} 

\subsection{\textcolor{black}{Fitting to simulation data}}
\textcolor{black}{In order to link the tidal deformability parameters $\lambda_{1,2}$ to quantities that can be measured from the gravitational wave signal, we make the following ansatz:
\begin{equation}
 \lambda_{1} = \left(\frac{\tilde \Lambda}{\Lambda_0}\right)^\alpha \left(\frac{M_{2}}{M_{1}}\right)^\beta,\quad \lambda_{2} = \left(\frac{\tilde \Lambda}{\Lambda_0}\right)^\alpha \left(\frac{M_{1}}{M_{2}}\right)^\beta
 \label{eq:lambda12}
\end{equation}
which encodes the fact that the lighter neutron star is more deformable than the heavier one. Here $\tilde \Lambda$ is the dimensionless tidal deformability parameter of the binary \citep{Raithel2018}. As a final tuning, we assume a floor disc mass of $M_\mathrm{d,min}=10^{-3}\msun$ as in \cite{Radice2018}.}

\begin{figure}
 \includegraphics[width=\columnwidth]{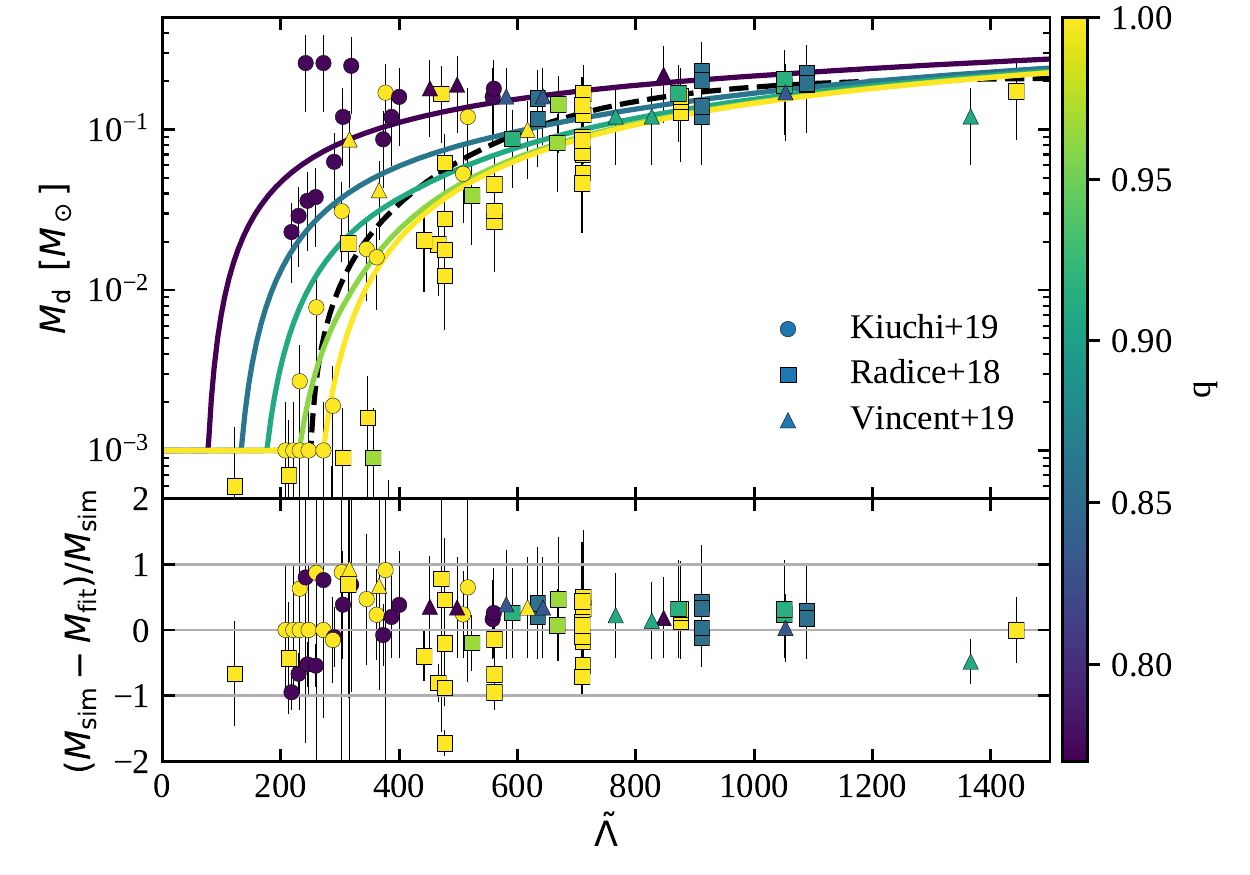}
 \caption{\textcolor{black}{Comparison between disc masses from numerical relativity simulations and the predictions of our fitting formula (Eqs~\ref{eq:Mdisk2}, \ref{eq:x_2} and \ref{eq:lambda12}). In both panels, datapoints show the disc masses reported in \citet[][squares]{Radice2018}, \citet[][circles]{Kiuchi2019} and \citet[][triangles]{Vincent2019}, as a function of the dimensionless tidal deformability parameter $\tilde \Lambda$ of the corresponding neutron star binary. The color of each marker shows the mass ratio $q$ of the binary, as coded in the colorbar on the right. In the upper panel, solid lines show the predictions of our fitting formula, assuming a representative total mass $M_1+M_2=3\,\mathrm{\msun}$. The black dashed line shows the fit from \cite{Radice2018} for comparison. The lower panel shows the relative residuals between the fitting formula (evaluated with the appropriate total mass for each binary) and the results from the simulation. More details in the text.}}
 \label{fig:Mdisk_fit}
\end{figure}

\textcolor{black}{The toy model has three free parameters, namely $\Lambda_0$, $\alpha$ and $\beta$. We determine these parameters by least-squares fitting the logarithm of the disc masses predicted by the model to the results of the numerical simulations presented in \citet{Radice2018}, \citet{Kiuchi2019} and \citet{Vincent2019}. We include all simulations reported in these works, despite some of them describe the same system but with differing setup (e.g.~different treatments of neutrino transport): this has the effect of including, in a crude way, the modelling uncertainty. We obtain the best fit values $\Lambda_0=245$, $\alpha=0.097$ and $\beta=0.241$. The result is shown in Figure~\ref{fig:Mdisk_fit}, where datapoints represent the disc masses as measured in the simulations, as a function of $\tilde \Lambda$. Squares, circles and triangles are data from \citet{Radice2018}, \citet{Kiuchi2019} and \citet{Vincent2019} respectively. The error bars represent the uncertainty in the disc mass as defined in \citet{Radice2018}. The upper panel shows representative curves $M_\mathrm{d}(\tilde\Lambda,q)$ from our fitting formula, for $q \in \left\lbrace0.77,0.86,0.91,0.96,1\right\rbrace$, assuming $M_1+M_2=3\,\mathrm{\msun}$ (for both datapoints and curves, the value of $q$ is color-coded according to the colorbar on the right). The lower panel shows the relative residuals between model and data (in this case, the appropriate total mass $M_1+M_2$ for each datapoint is used). The absolute relative residuals are below $0.5$ for $68\%$ of the simulations, and below $0.9$ for $90\%$ of them. Note also the close similarity between the equal-mass case (yellow line) and the fitting formula by \citet{Radice2018} (black dashed line, shown for comparison).}

\end{document}